\documentclass[journal,comsoc]{IEEEtran}
\usepackage[T1]{fontenc}

\ifCLASSINFOpdf
\else
\fi
\usepackage{amsmath}
\usepackage[cmintegrals]{newtxmath}
\usepackage{graphicx}
\usepackage{color}

\newcommand{\MYfooter}{\smash{\scriptsize
\hfil\parbox[t][\height][t]{\textwidth}{\centering
\copyright 2020 IEEE. Personal use of this material is permitted. Permission from IEEE must be obtained for all other uses, including reprinting/republishing this material for advertising or promotional purposes, collecting new collected works for resale or redistribution to servers or lists, or reuse of any copyrighted component of this work in other works. DOI: 10.1109/LCOMM.2020.2981070.}\hfil\hbox{}}}

\makeatletter

\def\ps@IEEEtitlepagestyle{%
\def\@oddfoot{\MYfooter}%
\def\@evenfoot{\MYfooter}}

\makeatother

\begin{document}
%
\title{Adaptive Multi-Receiver Coded Slotted ALOHA for Indoor Optical Wireless Communications}
%
%
%

\author{Dejan Vukobratovi\'{c},~\IEEEmembership{Member,~IEEE}, and Francisco J. Escribano,~\IEEEmembership{Senior Member,~IEEE}
\thanks{Dejan Vukobratovi\'{c} is with the Department of Power, Electronics and Communications Engineering, University of Novi Sad, 21000, Novi Sad, Serbia (e-mail: dejanv@uns.ac.rs).}%
\thanks{Francisco J. Escribano is with the Signal Theory and Communications Department, Universidad de Alcal\'{a}, 28805 Alcal\'{a} de Henares, Spain (email: francisco.escribano@uah.es).}
}

\maketitle

\begin{abstract}
In this paper, we design a novel high-throughput random access scheme for an indoor optical wireless communication (OWC) massive Internet of Things (IoT) scenario. Due to the large density of both IoT devices and OWC access points (APs), we base the proposed scheme on multi-receiver coded slotted ALOHA. In this scenario, collisions at APs are resolved by a centralized interference cancellation decoder that may exploit both spatial and temporal diversity. By applying adaptive control of each OWC AP field of view (FOV), the proposed system is able to dynamically adapt to different IoT device activation rates, in order to maintain a high total throughput. Using illustrative simulation results, we demonstrate the design methodology and performance possibilities of the proposed method.
\end{abstract}


%
\IEEEpeerreviewmaketitle

\section{Introduction}
\label{intro}
\IEEEPARstart{I}{ndoor} wireless connectivity is becoming further empowered by short-range optical wireless communications (OWC) \cite{Elgala2011}\cite{arnon2015visible}. Visible light communications (VLC) using light-emitting diodes (LEDs) may coexist with radio-frequency (RF) technologies while providing support for additional gigabit data rates as part of the emerging Li-Fi concept \cite{ayyash2015coexistence}\cite{dimitrov2015principles}, or used in a stand-alone deployments for use where RF communications are not desirable. Besides providing high data rates in the downlink, OWC can provide an efficient alternative for indoor Internet of Things (IoT) systems \cite{teli2018optical}\cite{haus2019enhancing}. For example, a variety of sensors could send their measurements using infra-red (IR) uplink transmission\footnote{For convenience of occupants, we assume the IoT devices emit the OWC signals outside the visible light spectrum, e.g., in the IR band.} to ceiling-mounted OWC access points (APs). Such OWC IoT model could be conveniently deployed in open-plan office spaces (e.g., sensors on employees' desks to monitor the environment), industrial halls or warehouses (e.g., sensors attached to shelves, production robots or unmanned vehicles), or LED-illuminated vertical farming (e.g., sensors that measure soil parameters).

An indoor OWC IoT system calls for a flexible and throughput-efficient uplink random access mechanism to accommodate sporadic and varying device activity. For multiple access, the IEEE 802.15.7 standard reuses carrier-sense multiple access with collision avoidance (CSMA/CA) \cite{ieee802_15_7}\cite{nobar2015comprehensive}, while other alternatives are considered for future OWC systems \cite{bawazir2018multiple}. However, for OWC IoT use cases, these solutions are suboptimal due to unpredictable user activity, and because of the directional nature of OWC links. In contrast, Slotted ALOHA (SA)-based approaches are identified as an efficient alternative \cite{zhao2016optimal}, which is the direction we pursue in this work.

An indoor OWC IoT scenario may contain a large number of IoT devices sporadically contending to send short packets to the backend through a number of ceiling-attached OWC APs. Due to upwards orientation, their signals may be detected at several OWC APs. We assume the OWC APs operate a mutually-synchronized framed SA (FSA) protocol, with IoT devices perfectly synchronized to slot timings. Following recent breakthroughs in throughput performance of FSA with interference cancellation (IC) decoding, we adopt the Coded SA (CSA) approach \cite{casini2007contention}\cite{Liva2011}. More precisely, we consider a multi-receiver (MR)-CSA model where each OWC AP forwards its received signals to a central processing entity, where IC decoding is applied in both time and spatial domains (i.e., across slots on each AP, and among slots at different APs \cite{jakovetic2015cooperative}). We examine the throughput of such a system in a realistic indoor setup as a function of user activity, and note that it highly varies depending on the OWC AP field of view (FOV). In other words, for different IoT device activity probabilities, there exist optimal FOV configurations that maximize the MR-CSA throughput. This leads to the proposal of an adaptive CSA solution where, at the beginning of each frame, the OWC IoT system estimates first the activity probability of the ensemble of devices, and then reconfigures the OWC APs FOV to maintain high-throughput operating points. A detailed system description and illustrative numerical results will provide details on system performance and appropriate design parameters.

The paper is accordingly organized as follows. Sec. II sets the indoor OWC IoT system model. In Sec. III, we provide details on the MR-CSA setup. An OWC IoT system that adaptively selects a throughput-maximizing FOV for the OWC APs is described in Sec. IV. Numerical results are presented in Sec. V, and the paper is concluded in the final section.

%
%
%
%
%
%

%
%

\section{System Model}
\label{model}

\begin{figure}[htbp]
\centering
\includegraphics[width=2.1in]{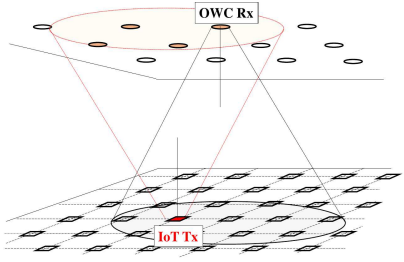}
\caption{System model: OWC IoT scenario. We can see the grid of IoT Tx on the floor plane, and of OWC Rx on the ceiling, along FOV examples.}
\label{Fig_1}
\end{figure}

\textbf{Indoor OWC IoT Setup:} We consider an indoor OWC IoT model where $N$ IoT devices contend to send data packets to $M$ OWC APs. We focus on uplink communication, with devices residing on a horizontal plane (e.g., floor or working plane), while ceiling-attached APs are placed on a regular square-grid. For simplicity, we assume devices are also placed on a regular square-grid, as shown in Fig. \ref{Fig_1}, which may reflect exactly or approximately a given real-world scenario. 

\textbf{OWC Signal Transmission and Reception:}  
OWC transmitters (Tx) at the IoT devices and OWC receivers (Rx) at the ceiling apply intensity modulation and direct detection, respectively, using optical signals with non-negative real values. This includes, e.g., on-off keying (OOK) or pulse position modulation (PPM), both included in IEEE 802.15.7 standard \cite{ieee802_15_7}. Their central radiation axes are assumed to be vertical, with OWC Tx's facing upwards and OWC Rx's downwards (Fig. \ref{Fig_1}). Assuming stationary devices and absence of moving obstacles, the signal at the receiver $j \in \{1,\cdots,M\}$ is:
\begin{equation}
 r_j\left(t\right)=\sum_{i \in \mathcal{U}_j} h_{ij} \otimes  s_i\left(t\right) + w_j\left(t\right),
\end{equation}
where $s_i\left(t\right)$ is the signal sent by the $i-$th device, $h_{ij}$ is the time-invariant optical channel gain\footnote{We adopt a Lambertian channel model for the line-of-sight (LOS) gain: 
\begin{equation}
\label{eq2}
       h_\textit{ij}  = 
        \begin{cases}
            \frac{A_r(m+1)}{2\pi{d_{ij}^2}}\mathrm{cos}^m(\phi_{ij}){g}(\psi_{ij}){T_s}(\psi_{ij})\mathrm{cos}(\psi_{ij}), & \text{ ${0}\leq{\psi_{ij}}\leq{\psi_c}$ } \\
            0, & \text{otherwise}
        \end{cases} \nonumber
\end{equation}
\begin{equation} 
\label{eq3}
m = \frac{-\ln{2}}{\mathrm{\ln{cos}}(\Phi_{1/2})}, \\ ~~~
\mathrm{g}(\psi)  = 
        \begin{cases}
            \frac{n_{r}^2}{\mathrm{sin^2}(\psi_c)}, & \text{for ${0}\leq{\psi_{i}}\leq{\psi_c}$ $$} \\
            0, & {0}\geq{\psi_c},
        \end{cases} \nonumber
\end{equation}
where $A_r$ is the physical area of the detector, $m$ is the Lambertian order of the light source, ${T_{s}}(\psi)$ is the gain of the optical filter, ${g}(\psi)$ is the gain of the optical concentrator, $\psi_c$ denotes the FOV of the receiver, $n_{r}$ is the refractive index of the concentrator, $\Phi_{1/2}$ is the semi-angle at half power of a light source, $d_{ij}$, $\phi_{ij}$ and $\psi_{ij}$ are the distance, radiation and incidence angle between the $i$-th Tx and the $j$-th Rx, respectively.} for the link between the $i$-th device and the $j$-th receiver, and $w_j\left(t\right)$ is the Gaussian noise. The sum is over the set $\mathcal{U}_j$ of all devices whose channel gains towards the $j$-th receiver satisfy $h_{ij} > 0$. Depending on the geometry, the signal from the $i$-th device may not reach the $j$-th receiver, resulting in $h_{ij}=0$. This establishes a model where each AP has a limited coverage, receiving data from devices located in its surrounding area, possibly overlapping with the area of neighboring receivers. As seen in Fig. \ref{Fig_1}, data from one device may be simultaneously received by more than one AP, a fundamental fact for the proposed MR-CSA scheme described next.
\begin{figure}[htbp]
\centering
\includegraphics[width=1.8in]{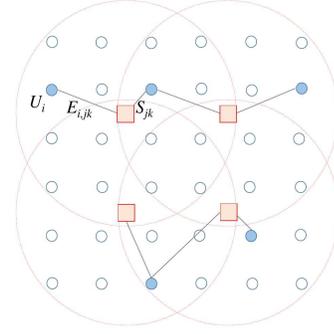}
\caption{CSA graph of a simple single-slot OWC IoT example. Circle-shaped nodes ($U_i$) represent device nodes (highlighting active devices as filled circles), while square-shaped nodes ($S_{jk}$) represent slot nodes at the receivers. The dashed circles represent the coverage area of each receiver. The edges ($E_{i,jk}$) represent a transmission from the corresponding device being received at the corresponding receiver during the given slot. In this example, $L=1$.}
\label{Fig_2}
\end{figure}

\section{Multi-Receiver CSA for Indoor OWC IoT}
\label{MR-CSA}

\begin{figure*}[htbp]
\centering
\includegraphics[width=5.2in]{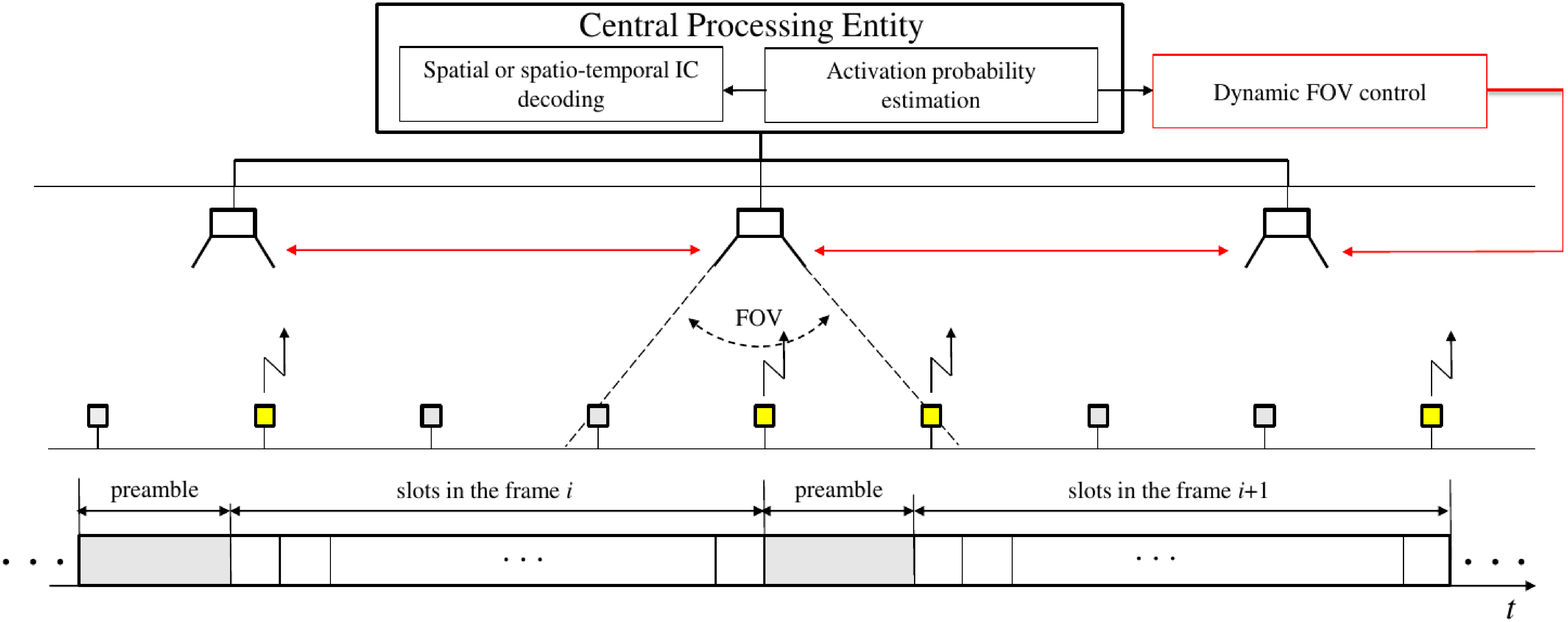}
\caption{Dynamic FOV adaptation. We depict the frame structure with the preamble, and a scheme of the central entity processing for $p_a$ calculation, FOV optimization and IC decoding. We can see the IoT Tx plane (below) and the OWC Rx plane (above). Active IoT devices are represented in brighter color.}
\label{Fig_3}
\end{figure*} 
\textbf{MR-CSA:} We consider a MR-CSA setup where the time is segmented into an uninterrupted sequence of frames, each one containing $L$ equal-duration slots. At the beginning of each frame, if the $i$-th device is active, it contends with other active ones to transmit a fixed-length data packet of $n_s$ symbols: $\mathbf{s}_i = \left( s_i \left[1\right], \cdots s_i\left[n_s\right]\right)$. We adopt a simple model in which every device is active with probability $p_{a}$ in every frame, regardless of the activity of other devices in that frame, and of its own activity in other frames. Within each frame, the active devices follow the baseline CSA method called irregular-repetition SA (IRSA) \cite{Liva2011}: every active device chooses to send $d$ replicas of the data packet in $d$-out-of-$L$ uniformly and randomly selected slots. Each device chooses this degree $d$ independently of other devices, and according to a common pre-defined degree distribution $\Omega(x)=\sum_{d=1}^{D}\Omega_d x^d$, where $\Omega_d$ is the probability that the device chooses $d$ replicas to send, $D \leq L$ is the maximum replication degree, and $\sum_{d=1}^{D}\Omega_d = 1$. We assume there are no distinctive delays in the transmission and reception across the network, making perfect synchronization within the whole network possible. The signal received at $j$-th receiver during slot $k=1, \cdots, L$ is:
\begin{equation}
 \mathbf{r}_{j,k}=\sum_{i \in \mathcal{U}_j} h_{ij} a_{i,k} \mathbf{s}_{i} + \mathbf{w}_{j,k},
\end{equation}
where $a_{i,k}=1$ if the $i$-th device sends its replica in the $k$-th slot, and $a_{i,k}=0$ otherwise, and $\mathbf{w}_{j,k}$ is the Gaussian noise vector for the $j$-th receiver at the $k$-th slot. We assume the channel gains $h_{ij}$ remain unchanged during a frame interval\footnote{Due to the stationarity of devices and receivers, and the absence of obstructions along the optical paths.}.

\textbf{MR-CSA Graph Model:} In the context of CSA, the previous setup can be conveniently represented using a graph-based model for the uplink transmission during a single frame. More precisely, the bipartite CSA graph $\mathcal{G} = \{\mathcal{U} \cup \mathcal{S}, \mathcal{E}\}$ consists in two classes of nodes: i) the set of $N$ device nodes $\mathcal{U}=\{U_i, 1 \leq i \leq N\}$, and ii) the set of $M \cdot L$ receiver slot nodes $\mathcal{S} = \{S_{jk}, 1 \leq j \leq M, 1 \leq k \leq L\}$, where $S_{jk}$ is the slot node that corresponds to the $k$-th slot at the $j$-th receiver. The sets $\mathcal{U}$ and $\mathcal{S}$ are connected by the set of edges $\mathcal{E} \subseteq \mathcal{U} \times \mathcal{S}$, where an edge $E_{i,jk}$ connects the device node $U_i$ with the slot node $S_{jk}$ if and only if $h_{ij}a_{i,k} > 0$, otherwise, $U_i$ and $S_{jk}$ are not connected. The CSA graph may be also characterized using the device node and the slot node degree distributions $\Lambda(x)=\sum_{d=1}^{U} \Lambda_d x^d$ and $P(x)=\sum_{d=1}^{S} P_d x^d$, respectively, where $\Lambda_d$ ($P_d$) is the fraction of device (slot) nodes of degree $d$, and $U$ ($S$) is the maximum device (slot) node degree. Therefore, $\Lambda_d$ ($P_d$) is the proportion of device (slot) nodes attached to exactly $d$ edges, and $U$ ($S$) is the maximum number of edges reaching the corresponding kind of node. An example of the CSA graph for a toy IoT OWC scenario is illustrated in Fig. \ref{Fig_2}. Following the MR-CSA approach \cite{jakovetic2015cooperative}, we consider the following two decoding scenarios.

\textbf{MR-CSA with Spatial Decoding:} We first consider a simple multi-receiver extension of SA. In this case, we have just one slot per frame, and the contention period reduces to a single slot. As in classical SA, if the $i$-th device is active within a given slot, it will send the data packet $\mathbf{s}_i$. In other words, we consider a special case of MR-CSA where $L=1$ and $\Omega(x)=x$. After each slot, all receivers forward their received signals $\mathbf{r}_j, 1 \leq j \leq M$ to the central processing entity. Then, \emph{the iterative spatial IC decoding} proceeds over the CSA graph $\mathcal{G} = \{\mathcal{U} \cup \mathcal{S}, \mathcal{E}\}$, assuming that both the CSA graph $\mathcal{G}$ and the corresponding channel gains $h_{ij}$ are known to the central decoder (for a static indoor scenario, these values can be determined in advance). In each iteration, the decoder processes the signals associated to slot nodes $\mathcal{S}$, and for each slot, it recognizes three scenarios: i) slot containing only noise (empty slot), ii) slot containing a single device signal (singleton), iii) slot containing two or more collided device signals (collision). We assume the decoder decodes without errors all singleton slots, and performs IC in spatial domain, i.e., removes the signal decoded at a singleton slot from all other slots that contain it. The decoding process continues iteratively until all device packets are decoded, or there are no new singleton slots at the beginning of an iteration. In the example of Fig. \ref{Fig_2}, the lower subgraph can be completely decoded in two iterations, while the upper subgraph will remain undecodable as there are no singleton slots.

\textbf{MR-CSA with Spatio-Temporal Decoding:} The performance of MR-CSA may be further improved by using spatio-temporal diversity, as proposed in \cite{jakovetic2015cooperative}. During the transmission phase, which now stretches across a frame of $L>1$ slots, we assume all active devices use the IRSA method and apply the same degree distribution $\Omega(x)$. In this case, the receivers forward their signals $\mathbf{r}_{j,k}$ to the central processing entity at the end of each frame. The central decoder is fully aware of the CSA graph $\mathcal{G}$ (i.e., the values of $a_{i,k}$) and the corresponding channel gains $h_{ij}$. The decoding uses the same IC processing of singleton slots, applied over the spatio-temporal CSA graph $\mathcal{G}$ expanded across the frame of $L$ slots.

\section{Adaptive FOV for OWC IoT}
\label{AdapFOV}

\textbf{Dynamic FOV Adaptation:} In the following, we consider the receiver FOV as a principal mechanism to dynamically optimize the throughput of the MR-CSA mechanism applied to the OWC IoT system. The proposed system is illustrated in Fig. \ref{Fig_3}. The receiver FOV $\psi_c$ may be adjusted using dynamic optical (lenses) or electro-mechanical (movable shades) systems attached to the receivers. By adjusting the FOV $\psi_c$, we directly affect the CSA graph $\mathcal{G}$ connectivity (i.e., the degree distributions $\Lambda(x)$ and $P(x)$), thereby affecting the decoding performance. Besides the receiver FOV, the structure of the CSA graph $\mathcal{G}$ depends on the set of active devices, which is random and induced by a single parameter: the activation probability $p_a$. Thus our goal is to find the optimal FOV $\psi_c^{*}$ in response to a given device activation probability $p_a$. More precisely, at the beginning of each frame, each active user sends a short preamble that enables the central processing entity to estimate the activation probability $p_a$. Then, the central system control adjusts the receiver FOVs $\psi_c$ to the optimal value maximizing the system throughput.

\textbf{Estimation of the activation probability:} In order to dynamically optimize the throughput by FOV adjustment, the system requires prior knowledge of the activation probability $p_a$. We leave out the details of such a scheme; however, for completeness, we provide some possible directions for this task. In the proposed MR-CSA scheme, we assume that each frame starts with a short period in which all active users simultaneously transmit a preamble. One option would be that all active devices send  the same short deterministic signal (e.g., a zero-one synch sequence). Based on the received preamble signal power, the known transmission power of the devices, and the known channel gains $h_{ij}$, the central processing entity could pose the activity probability estimation as a Bayesian inference problem and solve it using the maximum-a-posteriori probability method, or any of its computationally-efficient approximations. Moreover, with slightly longer preambles in the form of random-like binary ID sequences, the system could employ compressed sensing based methods to acquire not only the estimation of $p_a$ (i.e., the sparsity parameter), but also the active user identities \cite{boljanovic2017user}. For convenience, in the following, we assume the central system is able to recover a sufficiently good estimate of $p_a$ during the frame preamble.  

\section{Numerical results}
\label{simulation}

We provide here a variety of simulation results for the setups presented in the previous Sections. These results give useful insights about the trade-offs that have to be taken into account when deploying these systems for massive IoT. The basic setup parameters are the following:
\begin{itemize}
 \item An open-plan hall with $50$m length and $50$m width.
 \item A mesh of transmitting IoT devices placed on the same plane over equally-spaced grid points, separated $2$m along each direction (totalling $N=26\times26=676$ devices).
 \item A number $M$ of receiving OWC APs placed on the ceiling, located on the same plane over equally-spaced grid points.
 \item Both grids are regular and centered wrt the room plan.
 \item Seen from above, the OWC AP grid intersecting points lie in the middle of the squares of the IoT device grid.
 \item The distance (height) from the IoT device plane and the OWC AP plane is $3$m.
 \item The frame length $L$ and the degree distribution $\Omega\left(x\right)$ for spatio-temporal decoding are variable parameters.
\end{itemize}
For each parameter set, we have run a total of $10^5$ frames, sweeping different $p_a$ and $\psi_c$ values, so as to get experimental estimations of the average probability of packet recovery $p_{rec}$, defined as the average number of decoded packets divided by the total number of packets sent, and of the average throughput $R_{avg}$,  defined as the average recovery probability times the average fraction of active devices ($p_a\times N$), and divided by the total number of available slots per frame ($M \times L$).
\begin{figure}[htbp]
\centering
\includegraphics[width=8.6cm]{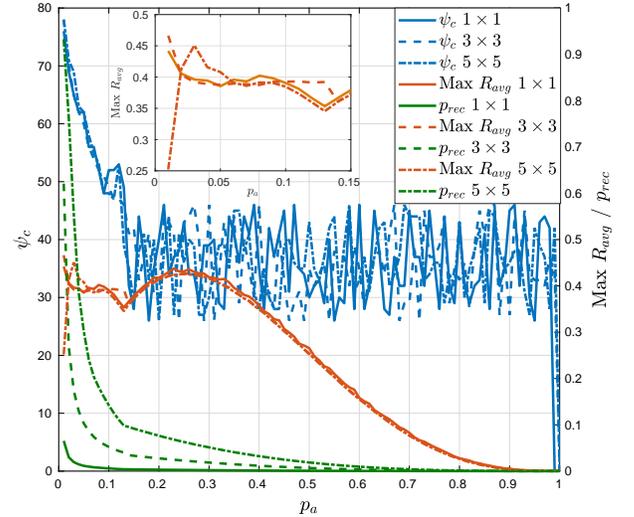}
\caption{Results for spatial decoding ($L=1$). Left y axis: FOV $\psi_c$ for maximum $R_{avg}$ as a function of the activation probability $p_a$. Right y axis: maximum $R_{avg}$ achievable as a function of $p_a$, and corresponding recovery probability $p_{rec}$.}
\label{Fig_4}
\end{figure}

Fig. \ref{Fig_4}  presents the results for MR-CSA with spatial decoding. We depict the FOV $\psi_c$ as a function of the activation probability $p_a$ that yields the maximum average throughput $R_{avg}$, for different quantitites of OWC APs: $1\times1$, $3\times3$ and $5\times5$. We also depict the value of maximum $R_{avg}$, and the corresponding value of $p_{rec}$. As it can be verified, the FOV follows quite similar trends regardless of the number of OWC APs: for small $p_a$, the FOV for maximum average throughput is high, and it falls very fast to a point where it keeps oscillating between arround $46^\textrm{o}$ and $26^\textrm{o}$. In this last situation, just the $4$ IoT devices closest to any of the OWC APs are covered, while the rest remain uncovered. The oscillations in this range are artifacts of the simulation, since any FOV among the mentioned values would lead to the same result.

We can also see that the evolution of the maximum average throughput $R_{avg}$ as a function of $p_a$ for the three situations is quite similar, while the average recovery probability $p_{rec}$ steadily improves as the number of OWC APs increases. The throughput keeps the same values from $p_a=0.14$, since from this value the coverage of each OWC AP shrinks to encompass just the closest $4$ IoT devices (this is just the equivalent of a pure SA per OWC AP). In this $p_a$ range, there is only an impact in $p_{rec}$, and there is no spatial diversity present. Nevertheless, when $p_a$ is small (lower than $0.05$), we can see that for $3\times3$ and $5\times5$ OWC APs a slightly higher maximum $R_{avg}$ is achieved with respect to the $1\times1$ case (zoomed area in the plot). In this situation, the spatial diversity is playing its intended role, since the FOV guarantees the presence of overlapping between coverage areas from adjacent OWC APs.

\begin{figure}[htbp]
\centering
\includegraphics[width=8.6cm]{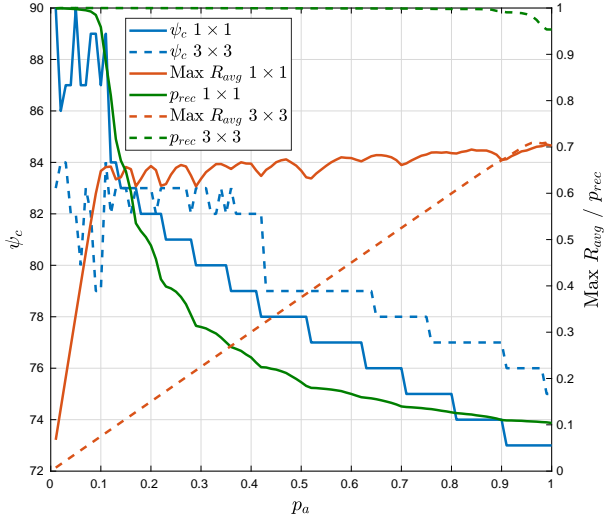}
\caption{Results for spatio-temporal decoding. Axis are as in Fig. \ref{Fig_4}. Plots are for $L=100$, and $\Omega\left(x\right)=0.498x^2+0.221x^3+0.038x^4+0.076x^5+0.040x^6+0.01x^7+0.09x^8+0.07x^9+0.03x^{11}+0.043x^{14}+0.08x^{15}+0.058x^{16}$.}
\label{Fig_5}
\end{figure}

 As seen, the effects of trying to exploit exclusively the spatial diversity for the setup chosen are limited, but the introduction of the temporal diversity (MR-CSA spatio-temporal decoding) leads to substantial improvements, as can be seen in Fig. \ref{Fig_5}. There we represent the results for a number of slots per frame $L=100$, a degree distribution $\Omega\left(x\right)$ optimized for the single-receiver RF environment \cite{Liva2011}, and a variable number of OWC APs. In the $1\times1$ case we have full coverage of the room for angles $\psi_c > 84^o$. After some initial oscillations, from $p_a > 0.1$, the coverage shrinks steadily, but the throughput is practically kept constant: the receiver is able to capture almost all the traffic from the devices under coverage because their number keeps decreasing, and $p_{rec}$ diminishes accordingly. In the $3\times3$ case, we can see how the spatial diversity dramatically changes the situation: almost all the offered traffic is recovered, excepting for $p_a$ close to $1$, and that is why the maximum value of $R_{avg}$ is an increasing straight line up to that point. It is to be noted that, in this case, there are no devices uncovered for the whole range, and there is always overlapping among coverage areas from adjacent OWC APs, since the minimum angle in the $3\times3$ case to have overlapping is $\psi_c=75^o$.

\begin{figure}[htpb]
\centering
\includegraphics[width=8.6cm]{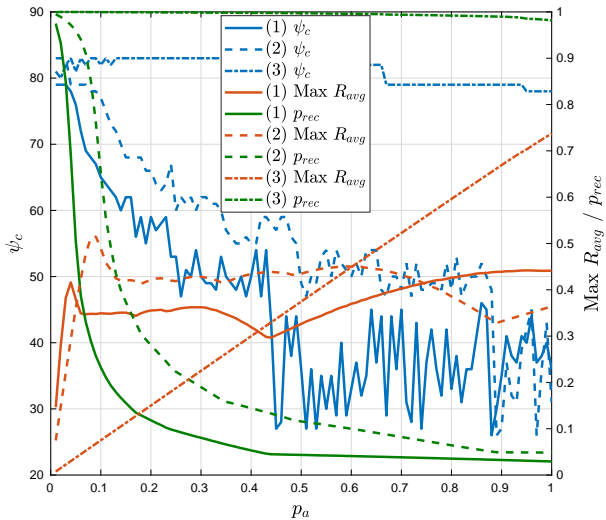}
\caption{Results for spatio-temporal decoding. Axis are as in Fig. \ref{Fig_4}. Plots are for $3\times3$ OWC APs and $\Omega\left(x\right)=x^2$. (1): $L=5$. (2): $L=10$. (3): $L=100$.}
\label{Fig_6}
\end{figure}

In Fig. \ref{Fig_6}, we have depicted several cases with $3\times3$ APs, and a degree distribution corresponding to the so-called contention resolution diversity SA (CRDSA) \cite{casini2007contention}, while the number of temporal slots are variable. As it may be seen, there is a progressive improvement as $L$ increases. In the case $L=5$, from $p_a=0.44$ the system falls into the regime with just $4$ IoT devices covered per OWC AP, while this only happens from $p_a=0.89$ in the case $L=10$. Moreover, the maximum value of $R_{avg}$ and the corresponding value of $p_{rec}$ improve accordingly. In both cases, nevertheless, there are uncovered IoT devices from low values of $p_a$ and on, but this does not happen for $L=100$. The results for $L=100$ are quite similar to the ones with the degree distribution in Fig. \ref{Fig_5}, and we have even better results as we approach $p_a=1$, which is in line with the results in \cite{jakovetic2015cooperative} that established the optimality of $\Omega(x)=x^2$ for high density of active users.

The previous results show how the spatio-temporal diversity can be successfully exploited in this context. It is clear that having large values of $L$ leads to the best results in throughput and recovery probability, but this also leads to poorer performance in data rate and delay, facts that have to be traded-off depending on the specific scenario.

\section{Conclusion}
In this work, we have presented a multi-receiver CSA setup for a massive OWC IoT scenario, with the possibility to exploit either spatial or spatio-temporal diversity. We have considered how the FOV of the OWC APs can be managed to get the highest possible throughput as a function of the device activation probability, and we have provided simulation results that give useful insights. We have illustrated that there is room for optimization in the different parameters that define the setup, as long as we introduce temporal diversity to complement the possibilities of spatial diversity. It has to be noted, however, that this comes at the cost of higher complexity, higher resources consumption, and higher processing delay. We are confident that subsequent works focusing on the resulting graph degree distributions and their possible optimization will lead to a better understanding, as well as to practical deployments of the solutions proposed in this article.


%
%
%
%

%

\ifCLASSOPTIONcaptionsoff
  \newpage
\fi

\end{document}